\documentclass[aps,twocolumn,preprintnumbers,nofootinbib,prl,superscriptaddress,10pt]{revtex4-1}

\makeatletter
\def\l@subsubsection#1#2{}
\def\l@subsubsubsection#1#2{}
\makeatother

\usepackage[utf8]{inputenc}
\usepackage[utf8]{inputenc}

\usepackage{soul}
\usepackage[normalem]{ulem}
\usepackage{enumitem}
\usepackage{graphicx,amssymb,amsmath,amsthm,amsfonts,epsfig,epsf}
\usepackage[usenames,dvipsnames,svgnames,table]{xcolor}
\usepackage{epstopdf}
\definecolor{darkred}{rgb}{0.5,0,0}
\usepackage{aas_macros}
\usepackage{bm}
\usepackage{dcolumn}
\usepackage{latexsym}
\usepackage{rotating}
\usepackage{longtable}

\usepackage{enumerate}
\usepackage{tensor,multirow}
\usepackage{url}
\usepackage[linktocpage]{hyperref}

\def\be{\begin{equation}}
\def\ee{\end{equation}}
\newcommand{\beq}{\begin{eqnarray}}
\newcommand{\eeq}{\end{eqnarray}}

\def\ba{\begin{align}}
\def\ea{\end{align}}

\newcommand\eq[1]{eq.~(\ref{#1})}

\begin{document}
\title{ Potential gravitational-wave signatures of quantum gravity}

\author{Ivan Agullo}
\affiliation{Department of Physics and Astronomy,Louisiana State University, Baton Rouge, LA 70803-4001, USA}
\author{Vitor Cardoso}
\affiliation{CENTRA, Departamento de F\'{\i}sica, Instituto Superior T\'ecnico -- IST, Universidade de Lisboa -- UL,
Avenida Rovisco Pais 1, 1049 Lisboa, Portugal}
\author{Adri\'an del Rio}
\affiliation{CENTRA, Departamento de F\'{\i}sica, Instituto Superior T\'ecnico -- IST, Universidade de Lisboa -- UL,
Avenida Rovisco Pais 1, 1049 Lisboa, Portugal}
\author{Michele Maggiore}
\affiliation{D\'epartement de Physique Th\'eorique and Center for Astroparticle Physics, Universit\'e de Gen\`eve, 24 quai Ansermet, CH–1211 Gen\`eve 4, Switzerland}
\author{Jorge Pullin}
\affiliation{Department of Physics and Astronomy,Louisiana State University, Baton Rouge, LA 70803-4001, USA}

\begin{abstract}
We show that gravitational-wave astronomy has the potential to inform us on quantum aspects of black holes. Based on Bekenstein's quantization, we find that black hole area discretization could impart observable imprints to the gravitational-wave signal from a pair of merging black holes, affecting their absorption properties during inspiral and their late-time relaxation after merger. In contrast with previous results, we find that black hole rotation, ubiquitous in astrophysics, improves our ability to probe  quantum effects. Our analysis shows that gravitational-wave echoes and suppressed tidal heating are signs of new physics from which the fundamental quantum of black hole area can be measured, and which are within reach of future detectors. Our results also highlight the need to derive predictions from specific quantum gravity proposals. 
\end{abstract}

\maketitle

\noindent{\bf{\em Introduction.}}
We are in the midst of a revolution in astrophysics and gravitation. The advent of gravitational-wave (GW) astronomy now allows a close scrutiny of 
 binaries of compact objects coalescing at close to the speed of light~\cite{Abbott:2016blz,Barack:2018yly}. In parallel, new techniques such as optical/infrared interferometry and radio 
large baseline interferometry have opened the possibility 
to measure matter in the close vicinity of black holes (BHs) with unprecedented accuracy~\cite{Abuter:2018drb,Abuter:2020dou,Akiyama:2019cqa}.
These new precision tools give us the ability to study strong-field gravity as never before, and make coalescing BH binaries the prime contender  to unravel  new physics beyond  general relativity (GR).

In this paper we explore the possibility that the GWs emitted in BH binary mergers carry information about the {\em quantum} properties of the BHs, and study the way this information can be extracted from observations.  
Quantum BHs are expected to have a discrete energy spectrum, and to behave in some respects like excited atoms. A general argument supporting this idea was originally formulated and explored in \cite{Bekenstein:1974jk} (see also \cite{Mukhanov:1986me,Bekenstein:1995ju,Kothawala:2008in}), based on the realization that the BH area $A$ behaves as an adiabatic invariant.
General arguments then give rise to a ``Bohr-Sommerfeld-like'' 
quantization of the area spectrum $A_N = \alpha \ell_p^2 N$, 
where $\ell_p=\sqrt{\hbar G/c^3}\sim 1.6\times 10^{-35}\, {\rm m}$ is the Planck length, $N$ a positive integer, and $\alpha\in \mathbb R$ is a phenomenological constant,
about which we will say more below. (We will use  units in which $c=G=1$.) The idea of BH area quantization has been materialized in theories of quantum gravity based on first principles, as for instance in loop quantum gravity \cite{Rovelli:1994ge,Rovelli:1996dv,Ashtekar:1997yu,Ashtekar:2000eq,Agullo:2008yv,Agullo:2010zz,FernandoBarbero:2009ai}, as we further discuss below. Similarly, the BH angular momentum is also expected to be quantized. Bekenstein and Mukhanov then concluded that BHs must have a discrete spectrum of mass, and worked out the consequences for the emission spectrum of BHs, i.e.\ for Hawking radiation \cite{Bekenstein:1995ju} {(see \cite{Hod:2015qfc} for the discussion of rotating BHs)}. We rather investigate the implications for  the {\em absorption} spectrum. 
Interestingly, although the area quantization takes place at the Planck scale, it can leave observable imprints on GWs~\cite{Foit:2016uxn,Cardoso:2019apo}. A simple calculation with a Schwarzschild BH serves to illustrate why. The area-mass relation $A=4 \pi (2M)^{2}$ implies that the mass $M$ can only change in discrete amounts
$\Delta M=\frac{\alpha \hbar}{32\pi}\frac{\Delta N}{M}\, .$
Thus, the frequencies that can be absorbed or emitted must also be quantized
\be \label{freq}
\omega = \frac{\left| \Delta M\right|}{\hbar}= \frac{ \alpha\,  \Delta N}{32 \pi M}\, . \ee
These frequencies  scale as $1/M$. Numerically, $\omega=c^3/(GM_{\odot})$ (where we have temporarily restored $c$ and $G$) corresponds to $f=\omega/(2\pi)\simeq 32.3\, {\rm kHz}$. Therefore, for  $M\simeq (10-50)~M_{\odot}$, as typical of the astrophysical BHs detected by  LIGO/Virgo, and  taking $\alpha=8\pi$ (see below) and  $\Delta N$ of order unity, the values of $f=\omega/(2\pi)$  are $O(10^2-10^3)$~Hz! 
Hence, astrophysical BHs act as ``magnifying lenses'', in the sense that they bring the Planck-scale discretization of the horizon within the realm of GW-observations.

The constant $\alpha$ determines the quantum of area of a BH. In Bekenstein's original proposal it takes the value $\alpha=8\pi$~\cite{Bekenstein:1974jk}.
Interestingly, this same value was obtained in \cite{Maggiore:2007nq} (elaborating on an earlier proposal in \cite{Hod:1998vk}), by 
modeling the quasinormal modes of a Schwarzschild BH, labeled by overtone number $n$, as  a collection of damped harmonic oscillators,
 $\ddot{\xi}_n+\gamma_{0,n}\dot{\xi}_n
+\omega_{0,n}^2\xi_n=0$. Demanding the behavior $\xi_n\propto e^{-\omega_{I,n}t \pm i\omega_{R,n} t}$,
where $(\omega_R+i\omega_I)_n$ is the complex frequency of the $n$-th quasi-normal mode, leads to the identification $\gamma_{0,n}/2=\omega_{I,n}$ and $\omega_{0,n}=\sqrt{\omega_{R,n}^2+\omega_{I,n}^2}$, whose large $n$ asymptotic behavior \cite{Motl03, MotlNeitze03} yields 
$\omega_{0,n}\simeq n/(4M)$. Now, if one interprets  the excitations of a quantum Schwarzschild BH by the collection of these damped harmonic modes, a transition between levels $n_1$ and $n_2$ can only emit or absorb the discrete frequencies $\omega\simeq \frac{ n_1-n_2}{4M}$. Two `miracles' happened here: first, transitions between QNMs reproduce the area linear quantization; and second, the constant $\alpha$ matches  the value obtained by Bekenstein with a completely different reasoning. Some other values of $\alpha$ derived in the literature with different arguments suffered from various inconsistencies, as discussed in \cite{Maggiore:2007nq}. Even if these arguments make  $\alpha=8\pi$ a preferred choice, we will keep $\alpha$ as a free parameter, and argue that it can be determined from observations. 

The arguments above neglect BH rotation~\cite{Foit:2016uxn,Cardoso:2019apo}. The consideration of BH spin has recently opened  a debate  in the literature. It was argued that astrophysical BHs may recover a continuum absorption spectrum due to spectral broadening enhanced by spin~\cite{Coates:2019bun}. If true, this would wash away any remnant of the underlying discrete BH energy spectrum.
However, a detailed analysis of the inclusion of spin is missing. We provide such study  here. As a result, we arrive at a different and novel framework, which does offer  observable predictions of BH area discreteness. We shall find that BH spin enriches the GW phenomenology in an unforeseen manner, increasing our ability to test the underlying hypothesis of BH area quantization.

\noindent{\bf{\em Problem statement}}. A complete analysis requires consideration of the full energy spectrum, including spinning BH states. The Kerr family of geometries is parameterized by the mass $M$ and angular momentum $J$
(and electric charge, which however is considered to be astrophysically irrelevant). In terms of the area, we have $M=\sqrt{\frac{A}{16\pi}+\frac{4\pi J^2}{A}}$.
Following Bekenstein's heuristic quantization, $A=\alpha \ell_p^2 N,\,J=\hbar j$, (where $j$ is a semi-integer number bounded by $0\leq j\leq \alpha N/8\pi$),  one finds
\beq
M_{N,j}=\sqrt{\hbar} \sqrt{\frac{\alpha N}{16\pi}+\frac{4\pi j^2}{N\alpha}} \, . \label{KerrM}
\eeq
The set of all $M_{N,j}$ for all allowed values of $N,j$ constitutes the energy spectrum of the quantum BH. Notice that it is much richer than the set of Schwarzschild states; in particular, it is highly non-uniform and this could invalidate on its own the results above for non-rotating BHs. In order to evaluate the potential physical consequences of  the discreteness of the  energy levels, we will address three questions: (i) What are the relevant energy transitions for the physical problem under consideration? I.e., are there selection rules that must be taken into account? (ii)  What is the  exact width $\Gamma$ of these energy levels? Do consecutive levels overlap, as suggested in~\cite{Coates:2019bun}? (iii) If there is no overlapping,
what are the expected  imprints in the interaction with GWs?

\noindent{\bf{\em Relevant BH transitions.}}
Since we are interested in the absorption of GWs in BH  binary mergers, we focus on interactions between an incident GW and a BH, and proceed similarly to what is done in atomic physics. Namely, radiative transitions of the BH can be studied with a Hamiltonian of the form $H(t)=H_{\rm BH}+H_{\rm int}(t)$, 
where $H_{\rm BH}$ denotes the (unperturbed) Hamiltonian of the quantum BH, and $H_{\rm int}$ describes the  interaction with the radiation field. An explicit expression for  these Hamiltonians would require a detailed understanding of the microscopic quantum theory. We adopt instead a phenomenological approach, and apply the familiar results of time-dependent perturbation theory in quantum mechanics. More precisely, we assume that $H_{\rm BH}$ has an orthonormal basis of eigenstates $\left|M\right>$ with eigenvalues $M_{N,j}$ given by  \eq{KerrM}. We then study transitions between different eigenstates caused by the interaction $H_{\rm int}$, which we treat perturbatively. 
Then, for the  interaction with a harmonic wave of frequency $\omega$,
the probability distribution is peaked around final energies $M_f=M_i \pm \hbar \omega$, where the plus/minus sign represents absorption/ emission. 
Notice that this reasoning requires focusing on transitions between different eigenvalues of the ADM mass, $M_{N,j}$---and not on transitions of the area quantum number $A_N$---since it is the quantity defined from the Hamiltonian of GR, $H_{BH}$, and to which the  frequencies relate to directly.

Gravitational perturbations around a BH background can be studied using the Newman-Penrose formalism~\cite{Teukolsky:1972my,Teukolsky:1973ha}. In this framework, the dynamics of relevant fluctuations is described by a wave-like equation for a master wavefunction describing the Weyl scalar $\psi_4$. One can use the isometries of the Kerr background and  decompose $\psi_4$ in modes $\psi_4\sim e^{-i\omega t} {_{-2}Y}_{\ell m}(\theta,\phi) R_{\omega \ell m}(r)/r^4$, characterized by the numbers $(\omega,\ell,m)$. The relative relevance of each mode 
is determined by the amplitude $R(r)$. The dominant mode
in GWs emission from interesting astrophysical systems (most notably quasi-circular inspirals, which comprehend nearly all of LIGO/Virgo events thus far) is the quadrupolar $(\ell=2,m=2)$ mode, and we will  focus our attention on it. 
Angular momentum conservation imposes then a selection rule, similar to the familiar ones in atomic physics.
This means that only energy levels differing in $\Delta j=2$ are relevant for this problem: 
\be
M_{N,j} \longrightarrow M_{N+\Delta N,j+2}\,.
\ee
These are the accessible energy levels of the BH when the $\ell=2, m=2$ GW mode impinges on it. This discussion answers question (i) raised above.

In view of this, the BH is unable to absorb the incident GW mode $(\omega,2,2)$ unless the frequency of the wave matches one of the characteristic frequencies,
$\hbar\omega_n = M_{N+n,j+2} - M_{N,j}$, with $n\equiv \Delta N$.
For large values of $N$, corresponding to macroscopic BHs, one obtains {from (\ref{KerrM})}
\beq
\hbar\omega_n
 =  \frac{\kappa \hbar}{8\pi }\alpha\, n +2\hbar\Omega_H+O(N^{-1}) \,,\label{m2transition2}
\eeq
where 
$\kappa=\frac{\sqrt{1-a^2}}{2M(1+\sqrt{1-a^2})}\,, $
$\Omega_H=\frac{a}{2M(1+\sqrt{1-a^2})}\,, $ 
are the surface gravity and angular velocity of the horizon, respectively, and $a\equiv J/M^2$ is the dimensionless BH angular momentum ($0\leq a\leq 1$). 
Notice that expression (\ref{m2transition2}) is in agreement with the first law of classical BH mechanics.

\noindent{\bf{\em The line-width.}}
%
The width $\Gamma$ of the energy levels can be written as the inverse of a decay rate, $\tau$, as $\Gamma=\hbar/\tau$. This timescale is intrinsically associated with the spontaneous decay of the BH energy states due to Hawking radiation, and it can be estimated as 
%
$\tau\equiv -\frac{\hbar \left<\omega\right>}{\dot M}\,, $ 
%
where $\left<\omega\right>$ denotes the average frequency over all possible decay channels, and $\dot M$ the power or luminosity (which is negative for the spontaneous Hawking decay). Both quantities can be computed using semiclassical arguments, following Page's calculations~\cite{Page:1976ki}
\beq
\dot M = - \sum_{\ell m} \int_0^{\infty}d\omega \, \hbar\omega \, \left<N_{\ell m}(\omega) \right>  ,\nonumber\\  
\left<\omega\right>= \frac{\sum_{\ell,m}\int_0^{\infty}d\omega \, \omega \, \left<N_{\ell m}(\omega) \right> }{\sum_{\ell,m}\int_0^{\infty}d\omega \,  \, \left<N_{\ell m}(\omega) \right>} \, , \nonumber
\eeq 
where $\left<N_{\ell m}(\omega) \right>$ is the   number expectation value. The luminosity can be written as $\dot M=-\hbar f(a)/M^2$, where the dimensionless function $f(a)$ depends only on the spin parameter $a$. We have computed $f(a)$, following Ref.~\cite{Page:1976ki} and considered emission into gravitons, photons, and the three families of neutrinos (the contribution from heavier fundamental particles is negligible). 
On the other hand, we find that  the combination  $M \left< \omega\right>$  is a dimensionless function of $a$, and it does not depend on the BH mass.
Gathering all the ingredients together, we find
\beq
\Gamma(M,a)=\frac{\hbar}{M} \frac{f(a)}{(M \left<\omega\right>)} \label{errorbar}\, .
\eeq
As expected, $\Gamma \propto M^{-1}$, confirming that heavier BHs are ``more stable''. 
Note also that we have assumed that the BH can emit any real frequency $\omega$ during the Hawking evaporation process. This is not  true if the energy levels are quantized, since only decay channels that end in a permissible value of the energy are allowed. Adding this restriction would make the BH more stable, and consequently would decrease $\Gamma$. Therefore, expression (\ref{errorbar}) overestimates $\Gamma$, and acts as an upper bound.

\noindent{\bf{\em No overlap of energy lines, and the critical $\alpha$.}}
\begin{figure}[h!]
\centering
\includegraphics[scale=0.4,trim=0cm 0cm 0cm 0cm]{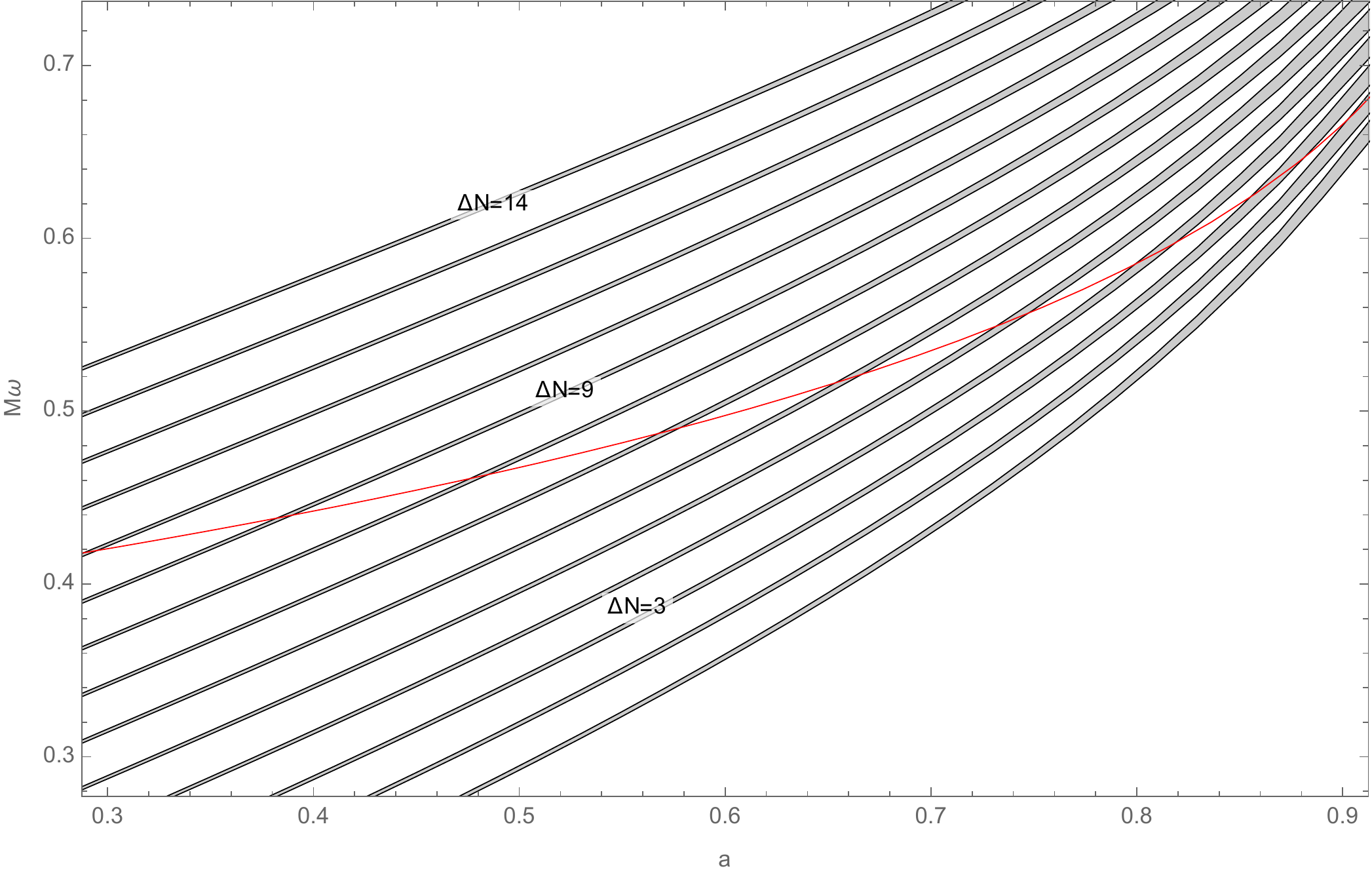}
\includegraphics[scale=0.4,trim=0cm 0cm 0cm 0cm]{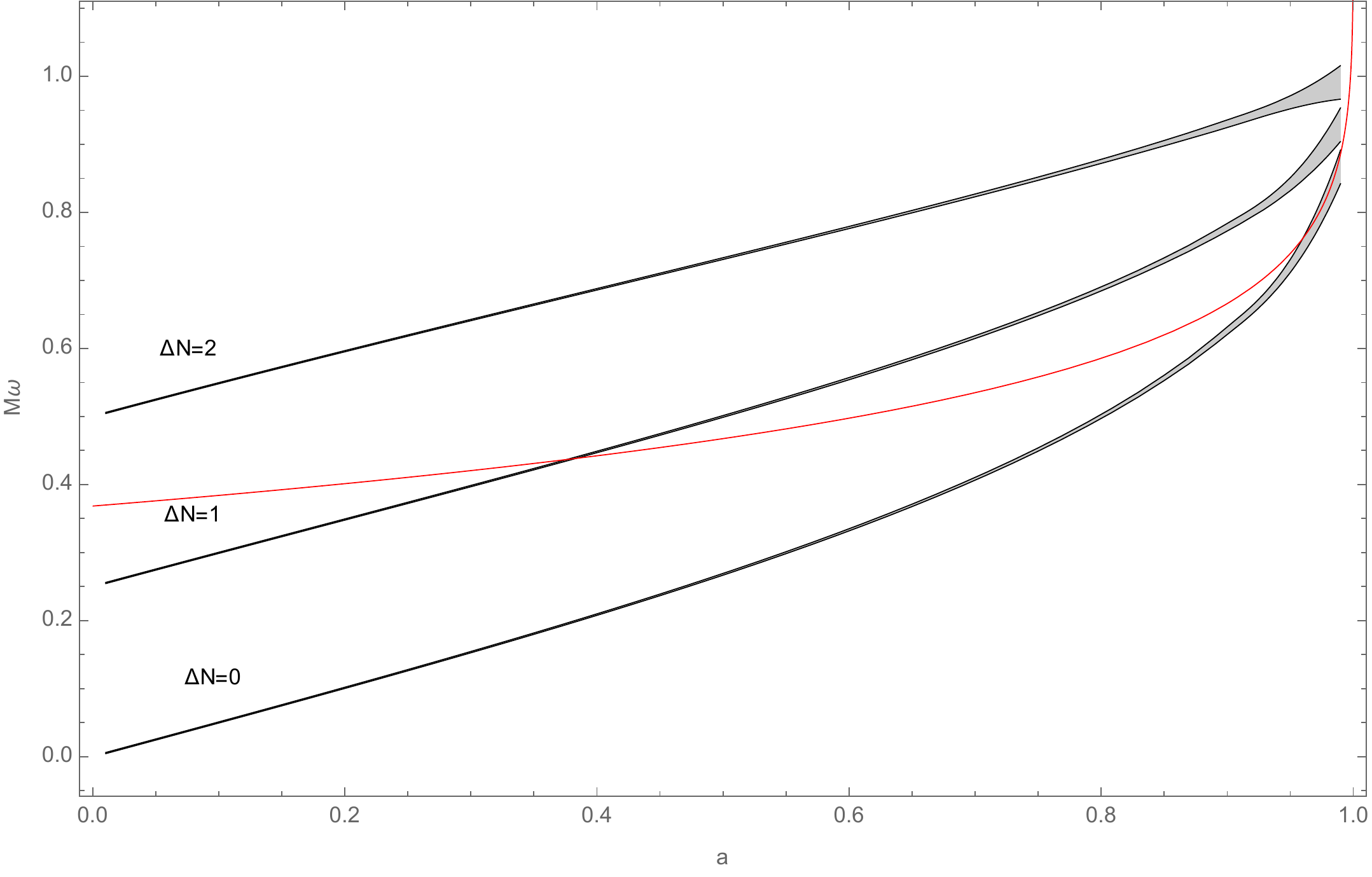}
\caption{BH absorption frequencies, line-widths and QNMs of spinning BHs. Black lines represent the frequencies $\omega_n$---defined in Eq.\ (\ref{m2transition2})---corresponding to the energy transitions of Kerr BHs characterized by $\Delta N$ and $\Delta j=2$, for different values of $\Delta N$, as a function of the rotation parameter $a$. The thickness of the black lines measures the width $\Gamma$ of the spectral lines. The upper panel shows these lines for $\alpha=4\log{2}$, while $\alpha=8\pi$ is shown in the bottom panel. The (real part of the) frequency $M\omega_{022}(a)$ of the dominant QNM is shown in red. The dominant QNM will be absorbed by the BH only for values of $a$ for which the red curve intersects one of the black lines.}
\label{fig:linewidth}
\end{figure}
We are now in a position to compare the BH energy levels with their widths.  Figure~\ref{fig:linewidth} shows these quantities,  as a function of the spin parameter $a$ and for two representative values of $\alpha$. This figure contains two  messages. On the one hand, the energy of the first  level, which corresponds to $n\equiv \Delta N=0$ in (\ref{m2transition2}) and is independent of $\alpha$, is larger than the width $\Gamma$ for all values of $a$, except for $a$ very close to 0 where the energy of this level vanishes. This implies that there is a minimum absorption frequency  $\omega =2\hbar\Omega_H$ for  the GW mode $(\omega,2,2)$. On the other hand, there is no overlap of {the other} spectral lines except for extremely high rotation. 
More precisely, the ratio between the width $\Gamma$ and the energy of  consecutive  levels, $R(a)\equiv\Gamma/[\hbar (w_n-w_{n-1})]$, depends on both $\alpha$ and $a$---the $M$ dependence cancels out. {We obtain that the critical value of $\alpha(a)$ below which  there is overlapping of the energy levels ($R(a)\geq 1$) is}
\be
\alpha_{\rm crit}(a)=0.0842+0.2605a^2+0.0320e^{5.3422a^3}\, ,
\ee
accurate to within $2\%$ for $a<0.9$. %
As an example, consider a BH binary of two non-spinning BHs. The remnant's spin was found through numerical relativity simulations to be $a \approx 0.7$~\cite{Buonanno:2006ui}.  This value is in good agreement  with
 the spin of the remnant BH of a large fraction of the observed mergers~\cite{LIGOScientific:2018mvr}. For this value of $a$ we obtain $\alpha_{\rm crit}=0.415$, which is one order of magnitude {\em below} $4 \log{2}$, the smallest value of $\alpha$ considered in the literature, and hence there is no overlap.   

 {We remark that these conclusions hold for the BH {\it absorption} spectrum in the coalescence of a binary  system, due to the existence of  selection rules imposed by the particular GWs generated in the process. 
However, for the spontaneous  (Hawking) {\it emission} spectrum any mode $(\ell,m)$ can be emitted, and  all levels in the Bekenstein  energy spectrum (\ref{KerrM}) are now accesible. The full spectrum is complex and  highly irregular  (in particular, the closest energy levels of $M_{N,j}$ are not simply $M_{N\pm1,j}$, but   involve  an energy-dependent change in $N$ and $j$). Using a computer,  we observe that the existence of many more accessible levels  partially recovers the original continuous Hawking  spectrum, giving support to previous ideas \cite{Hod:2015qfc}.  The emission spectrum is not the main goal of this paper however, and it deserves further analysis.} 

The analysis of this and the previous section answers question (ii), and we now focus on  (iii).

\noindent{\bf{\em Gravitational-wave echoes.}}
Observational consequences of BH area quantization could include a distorted ringdown signal, or most likely the presence of late-time echoes~\cite{Cardoso:2019apo}. The ringdown signal from the merger of a BH binary, as computed from GR, is described very well by the QNM of the final Kerr BH. For the dominant quadrupole mode, $\ell=2$, the frequencies of the QNM for a Kerr BH can be found in Table VIII of \cite{Berti:2005ys} for all $m$, and for the first overtones $n=0,1,2$. For example, the fundamental mode, $n=0, \, l=m=2$, is well described by the expression
$M{\rm Re}\, \omega_{022}\simeq 1.5251-1.1568(1-a)^{0.1292}\,.$ 
The QNM modes, ``localized'' at the BH photonsphere, are excited during the  merger and start propagating. A fraction of the energy in these modes moves outwards to GW detectors, but a significant fraction is directed inwards towards the horizon, where it is absorbed according to GR~\cite{Cardoso:2016rao,Cardoso:2016oxy,Cardoso:2019rvt}. However, when quantum effects are considered, absorption takes place only if the oscillation frequency of the QNM matches one of the transition lines of the Bekenstein spectrum, i.e., if
\be \label{freqQNM}
{\rm Re}\, \omega_{022} \in \left[\omega_{n}- \frac{\Gamma }{2\hbar}, \omega_{n}+ \frac{\Gamma }{2\hbar} \right]
\ee
where $\omega_{n}$ is given by Eq.\ (\ref{m2transition2}). Figure~\ref{fig:linewidth} compares $\omega_{022}$ and  $\omega_{n}$, for $\alpha=4\log{2}$ and $\alpha=8\pi$. We observe that for these values of $\alpha$ the absorption of the dominant QNM will be suppressed, except for a set of values  of $a$, and consequently echoes are expected for values of $a$ between consecutive intersections of the red line and black lines in Fig.~\ref{fig:linewidth}. This, in turns, offers a way of determining $\alpha$: if echoes are observed for binary mergers sampling a large enough range of $a$, their frequencies will provide a direct measurement of $\alpha$. 

The ability of measuring $\alpha$ is of obvious interest for fundamental theories of quantum gravity. For instance, in  loop quantum gravity (LQG) the canonical BH area spectrum is not equally spaced \cite{Rovelli:1994ge,Rovelli:1996dv,Ashtekar:1997yu,Ashtekar:2000eq,Agullo:2008yv,Agullo:2010zz}, and the gap between consecutive values decreases exponentially with the area \cite{Barreira:1996dt,Barbero:2017jxa}. One finds an almost continuum area spectrum for macroscopic BHs, and therefore no echoes are expected in GW observations. However, an alternative way of defining the operator associated with the area of a BH in LQG has been recently proposed in~\cite{FernandoBarbero:2009ai}, in which the BH area spectrum has indeed equally spaced eigenvalues, in agreement with Bekenstein's  proposal, and it predicts $\alpha=4 \log 3$. In such a scenario, BH mergers would generate GW echoes. Our analysis shows that forthcoming observations will be able to discriminate between these two possibilities.

The ability to detect echoes in GW signals depends on how much energy is converted from the main 
burst into echoes, and on the ability to produce faithful templates. Following the initial suggestion in
\cite{Abedi:2016hgu},
fully Bayesian searches for echos in the LIGO/Virgo data, tied to  phenomenological families of echo waveforms, do not find evidence for echoes and rule out echos amplitude as large as $0.1-0.2$ relative to the original signal 
peak ~\cite{Westerweck:2017hus,Nielsen:2018lkf,Lo:2018sep,Uchikata:2019frs}. These constraints will improve significantly in the near future with ground-based 3G detectors such as the Einstein Telescope (ET)~\cite{Maggiore:2019uih} and the planned space mission LISA~\cite{Audley:2017drz}.

We remark that  QNM perturbations were treated here as  wave fronts. Their pulse character and other sources of uncertainty in the energy of individual quanta may add corrections to the intensity of  echoes. However, a quantitative analysis would require  inputs on the microscopic interaction between BHs and radiation.

\noindent{\bf{\em Tidal heating.}}
%
As we showed, finite but small values of $\alpha$ will lead to peculiar echoes in the GW signal.  
In addition to the effects on the ringdown phase, an imprint of BH area quantization can  also be present in the early inspiral, where the GW frequency is even smaller than the one during the ringdown phase. 
Classically, individual components of a BH binary absorb GWs at a rate which, although a small fraction of the rate of radiation to infinity, is not negligible~\cite{Poisson:1994yf}.
These waves produce tidal forces that act on the bodies causing distortions of their event horizons \cite{Goswami:2019fyk}. But as a BH rotates under this bulge,  its rotational energy is dissipated gravitationally \cite{Hartle:1973zz}, and it is transferred to the orbital motion of the binary. This phenomenon is known as tidal heating.

Consider a binary made of spinning BHs. Due to the energy gap of the Bekenstein's energy spectrum, the absorption of low frequency GWs impinging on each individual BH is now highly suppressed, and this causes a change in the binary evolution with respect to the prediction of classical GR \cite{Hughes:2001jr}. 
In particular, tidal heating affects the GW as a 2.5 PN ($\times \log v$, with $v$ the orbital velocity) correction to the GW phase of spinning binaries, relative to the leading term~\cite{Maselli:2017cmm}. A promising strategy is therefore to parametrize the waveform with an absorption parameter $\gamma$ multiplying the 2.5 PN$\times \log v$ GR term~(see Ref.~\cite{Maselli:2017cmm} for details).
For classical BHs $\gamma=1$, and one recovers the standard GW phase of quasi-circular BH binaries. However, when area quantization is taken into account, and for highly spinning BH binaries, the energy gap $2\hbar \Omega_H$ in Eq.~\eqref{m2transition2} suggests that absorption is highly suppressed for the entire duration of the inspiral,
and  one expects $\gamma\ll 1$. Can we use GW signals to discriminate between $\gamma=0,1$? This study was done recently in the context of exotic compact objects~\cite{Maselli:2017cmm,Cardoso:2019rvt,Datta:2019euh,Datta:2019epe,Datta:2020gem}, and the conclusions can be extended to our setup: advanced detectors such as LISA and ET have a strong potential to discriminate between absorption or no-absorption at the horizon.

\noindent{\bf{\em Discussion.}}
Testing Planck-scale physics with kHz interferometers is a mind-blowing prospect; as now usual in BH physics, this possibility seems to be open due to the ``holographic'' properties of BHs, in the sense that {their} mass squared is proportional to {their} surface area.
We have shown that the merger of two BHs may not generate GWs with the appropriate frequency so as to excite transitions in quantum states, leading to peculiar features in GW signals, most notably echoes and modified GW phase at $2.5$ PN order in the ringdown and inspiral phases, respectively. We note however, that the number of gravitons hitting each BH is very large. The use of the quadrupole formula~\cite{MaggioreBook} yields $\dot{N}\sim7\times 10^{74}\left(\frac{M}{10M_{\odot}}\right)^{9/2}$ gravitons emitted per second. Of these, a fraction $\sim v^8$, with $v$ the orbital velocity, goes down the horizon (at least in an EMRI)~\cite{Poisson:1994yf}. This is equivalent to $\sim 10^{64}$ gravitons crossing the horizon of a stellar-mass BH, per second. 
 Multi-graviton effects may become relevant~\cite{doi:10.1002/andp.200910358,PhysRevA.4.1896} and deserve further study.

The analysis done in this work rests on Bekenstein-Mukhanov semi-heuristic arguments on quantum BHs. A more {accurate} description requires an understanding of the fundamental BHs degrees of freedom, and the way they interact with the radiation field. That these aspects of BHs are within the reach of forthcoming observations should encourage {advances} in  frameworks of quantum gravity to develop concrete predictions. 

{After submission of this work, it has been suggested that quantum effects may also have an impact on the tidal deformability of black holes, with testable observation consequences~\cite{Brustein:2020}. This effect adds to the one described here and is in fact consistent with our approach.} {Other approaches that do not involve area quantization have reached different conclusions \cite{Goldberger:2019sya}}.

\noindent{\bf{\em Acknowledgments.}}
%
We are indebted to A. Coates, S. Volkel,  K. Kokkotas, and M. Van de Meent   for useful comments and correspondence.
V. C. would like to thank Waseda University for warm hospitality and support. 
V. C. and A.D.R. acknowledges financial support provided under the European Union's H2020 ERC 
Consolidator Grant ``Matter and strong-field gravity: New frontiers in Einstein's 
theory'' grant agreement no. MaGRaTh--646597. 
This project has received funding from the European Union's Horizon 2020 research and innovation 
programme under the Marie Sklodowska-Curie grant agreement No 690904.
We thank FCT for financial support through Project~No.~UIDB/00099/2020 and through grant PTDC/MAT-APL/30043/2017.
The authors would like to acknowledge networking support by the GWverse COST Action 
CA16104, ``Black holes, gravitational waves and fundamental physics.'' The work of
MM is supported by the  Swiss National Science Foundation and  by the SwissMap National Center for Competence in Research. 
IA  is supported by the NSF CAREER grant PHY-1552603 and by the Hearne Institute for Theoretical Physics. 
JP is supported by grant NSF-1903799, by the Hearne Institute for Theoretical Physics and CCT-LSU.

\bibliographystyle{utphys}
\bibliography{References}

\end{document}